\documentclass[aps,amsmath,amssymb,amsfonts,twocolumn]{revtex4-2}

\usepackage{graphicx}
\usepackage[colorlinks=true,linkcolor=blue,citecolor=blue, urlcolor=blue]{hyperref}
\usepackage{blindtext}

\usepackage{sidecap}
\usepackage{lineno}
\usepackage{setspace}
\AtBeginDocument
{
	\addtolength\abovedisplayskip{0.20\baselineskip}%
	\addtolength\belowdisplayskip{0.20\baselineskip}%
}

\begin{document}

\title{Conformational isomerization dynamics in solvent violates both the Stokes-Einstein relation and Kramers' theory}

\author{Benjamin A. Dalton}
\affiliation{Freie Universit\"at Berlin, Fachbereich Physik, 14195 Berlin, Germany}

\author{Henrik Kiefer}
\affiliation{Freie Universit\"at Berlin, Fachbereich Physik, 14195 Berlin, Germany}

\author{Roland R. Netz}
\affiliation{Freie Universit\"at Berlin, Fachbereich Physik, 14195 Berlin, Germany}

\date{\today}

\begin{abstract}
Molecular isomerization kinetics in liquid solvents are determined by a complex interplay between the friction acting on a rotating dihedral due to interactions with the solvent, internal dissipation effects (also known as internal friction), the viscosity of the solvent, and the free energy profile over which a dihedral rotates. Currently, it is not understood how these quantities are related at the molecular scale. Here, we combine molecular dynamics simulations of isomerizing n-alkane chains and dipeptide molecules in mixed water-glycerol solvents with memory-kernel extraction techniques to directly evaluate the frequency-dependent friction acting on a rotating dihedral. We extract the friction and isomerization times over a range of glycerol concentrations and accurately evaluate the relationships between solvent viscosity, isomerization kinetics, and dihedral friction. We show that the total friction acting on a rotating dihedral does not scale linearly with solvent viscosity, thus violating the Stokes-Einstein relation. Additionally, we demonstrate that the kinetics of isomerization are significantly faster compared to the Kramers prediction in the overdamped limit. We suggest that isomerization kinetics are determined by the multi-time-scale friction coupling between a rotating dihedral and its solvent environment, which results in non-Markovian kinetic speed-up effects.
\end{abstract}

% , which have recently been shown to characterize the folding kinetics for fast-folding proteins

\maketitle

\section*{Introduction}

Molecular conformation transition rates, such as the folding rates of proteins, are influenced by interactions between the molecule and its solvent environment, as well as by intra-molecular interactions within the molecule itself. In experimental settings, the molecular conformation dynamics of a solute molecule can be modulated by altering the viscosity of the solvating medium. One way to achieve this is to incorporate viscogenic agents like glucose, sucrose, or glycerol into the medium. By doing so, one can generate a solvent viscosity $\eta$ that is far greater than that of pure water. This method has been widely used in the field of protein folding and has played a critical role in uncovering the importance of internal friction effects \cite{Ansari_1992, Jas_2001, Hagen_2010, Beece_1980, Doster_1983, Jacob_1999, Soranno_2012, Borgia_2012}, whereby the viscosity scaling of the folding time $\tau$ was typically written as $\tau =\alpha\eta^{\beta} + \varepsilon$ such that it was argued that $\beta = 1$ and $\varepsilon = 0$ in the absence of internal friction and that either $\beta < 1$ or $\varepsilon > 0$ when internal friction effects are present. Applying similar methodologies, all-atom simulations have also been used to elucidate the molecular mechanisms of internal friction \cite{Einert_2011, Schulz_2012, Echeverria_2014, DeSancho_2014,  Zheng_2015, Daldrop_2018}. The supposed linear relation between $\eta$ and $\tau$ (i.e. $\beta = 1$) with $\varepsilon = 0$ is actually founded on the combination of two more fundamental relations, which have been impossible to check separately. According to the Stokes-Einstein relation, the friction $\gamma$ acting on a molecule that moves through a solvent of viscosity $\eta$ satisfies $\gamma \sim \eta$, with a pre-factor that incorporates information about the molecule's geometry. For sufficiently over-damped systems, Kramers' theory \cite{Kramers_1940} tells us that the average time $\tau$ to undergo a state transition by overcoming an energy barrier satisfies $\tau \sim \gamma$, with a pre-factor that incorporates information about the energy barrier. Therefore, the linear relation between $\tau$ and $\eta$ is indirectly mediated by the friction acting on the reconfiguring molecule, with deviations from linearity actually indicating violations of either the Stokes-Einstein relation, the overdamped Kramers relation, or both. While it is relatively commonplace to measure transition times and solvent viscosities, both in experiments and in simulations, a direct evaluation of the friction acting on some collective reaction coordinate is far more complicated. Therefore, a direct verification of the Stokes-Einstein relation and the overdamped Kramers relation for molecular isomerization in complex viscogenic solvents has so far not been possible.
\begin{figure*}[!t]
\includegraphics[scale=1.0]{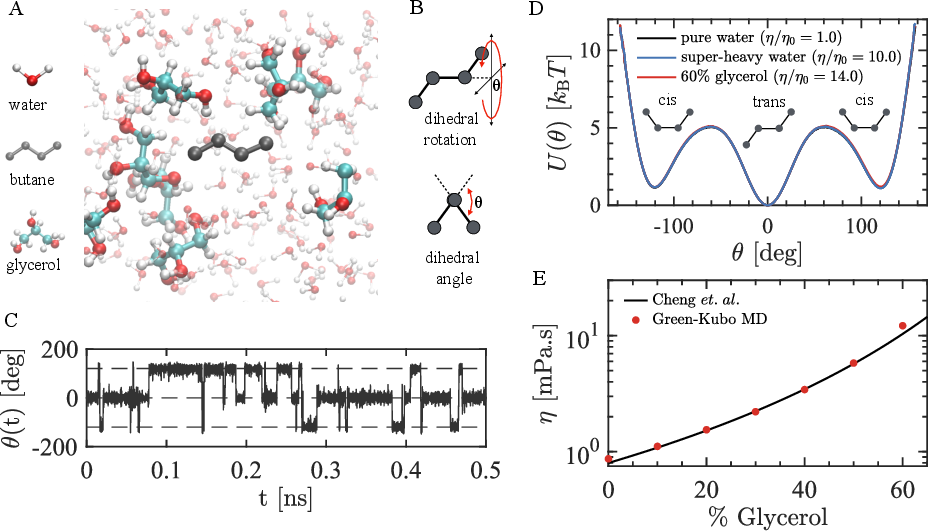}
\caption{Simulation of butane in explicit solvents. (A) Simulation snapshot of a butane molecule dissolved in a water-glycerol mixture at 20$\%$ mass-fraction glycerol. (B) Schematic indicating the butane dihedral. The dihedral angle $\theta$ is subtended by the intercept of the two planes formed by the four carbon-hydrogen groups. (C) A 0.5 ns trajectory segment of the butane dihedral angle in a standard pure water solvent. The dashed lines show $\theta=\pm120$ deg and $\theta=0$ deg. (D) Free energy profiles for the butane dihedral extracted from MD simulations for three different solvent conditions. The schematics indicate the configurations of the cis and trans states, which occupy the various energy minima. (E) Viscosity for water-glycerol mixtures, calculated using the Green-Kubo method, plotted as a function of glycerol mass fraction. Results are compared to an experimental empirical curve by Cheng et al. \cite{Cheng_2008}. \label{Figure_Methods}}
\end{figure*}

To evaluate friction, in the past one typically relied on indirect methods using memoryless reaction-rate theory \cite{Best_2006, Best_2010, Hinczewski_2010, Chung_2015}. In this paper, we utilize recent non-Markovian memory kernel extraction methods \cite{Daldrop_2018, Kowalik_2019, Mitterwallner_2020, Ayaz_2021, Brunig_2022a, Brunig_2022d, Dalton_2023} to directly evaluate the frequency-dependent friction acting on a rotating dihedral. Memory extraction methods enable a direct evaluation of the friction acting on any arbitrary reaction coordinate by mapping the time series evolution of that reaction coordinate onto a generalized Langevin equation (GLE) \cite{Zwanzig_1961, Mori_1965}. These methods are remarkably general and have been applied recently to butane isomerization \cite{Daldrop_2018}, cell migration \cite{Mitterwallner_2020}, the vibrational spectra of water molecules \cite{Brunig_2022a}, pair reactions in water \cite{Brunig_2022d}, the dynamics of small polypeptide chains \cite{Ayaz_2021}, and, most recently, to the folding dynamics of a diverse set of fast-folding proteins \cite{Dalton_2023}. To investigate the relationship between dihedral friction, solvent viscosity, and isomerization kinetics, we simulate four n-alkane chains: n-butane, n-hexane, n-octane, and n-decane (hereafter, we omit the \textit{n} prefix), and two amino acid residues: alanine and phenylalanine, both with NMA C-terminal capping and ACE at N-terminal capping, using molecular dynamics (MD) simulations with explicit solvents. As a viscogenic agent, we mix water and glycerol, and we vary the concentration of glycerol to change the solvent viscosity. We compare our results to an idealized system where the viscosity of a pure water solvent is modified by scaling the mass of the water molecules. We observe that the dependence of the friction governing the dihedral rotation on the solvent viscosity is the same, regardless of whether we vary the glycerol concentration or the mass of the water molecules. However, the scaling is significantly sublinear in both cases, showing that dihedral isomerization strongly violates the Stokes-Einstein relation. Interestingly, the mean first-passage times $\tau_{\rm{MFP}}$ for dihedral isomerization exhibits dramatically different viscosity scaling depending on whether we vary the glycerol concentration or the water mass. This disagreement is most extreme for the smallest solute butane and reduces as a function of solute size. When we evaluate the dependence of $\tau_{\rm{MFP}}$ on $\gamma$ we find that the linear scaling $\tau_{\rm{MFP}} \sim \gamma$ does not hold and that $\tau_{\rm{MFP}}$ is significantly reduced compared to the Kramers prediction in the overdamped limit. This dramatic acceleration of reconfiguration kinetics was recently reported for a set of extensive fast-folding protein simulations \cite{Dalton_2023}, where it was shown that many proteins fold and unfold in a memory-induced barrier-crossing speed-up regime \cite{Kappler_2018}. The same non-Markovian mechanism also applies to dihedral isomerization kinetics. Overall, our investigation reveals the full complexity of the relationships between friction, viscosity, and reaction kinetics at the molecular scale.

\section*{Results and discussion}

\noindent \textbf{Viscosity dependence of butane isomerization kinetics. }To begin, we study the viscosity dependence of the dihedral dynamics of butane, which is the smallest molecule to exhibit distinct isomeric states and has been used as a model system for many classic studies in the statistical mechanics of dihedral barrier-crossing processes \cite{Chandler_1978, Rebertus_1979, Montgomery_1979, Rosenberg_1980, Levy_1979}. We modify the viscosity by either varying the concentration of glycerol in a mixed water-glycerol solvent (denoted as w/gly throughout) or by scaling the mass of the water molecules in a pure water solvent (referred to as super-heavy water and denoted as $\Delta m_{\text{w}}$ throughout). In the case of the super-heavy water, we uniformly change the mass of the water molecules such that the viscosity $\eta$ scales as $\eta/\eta_0=\sqrt{{m}/{m}_0}$, where ${m}_0$ and $\eta_0$ are the mass and viscosity of neat water, and $m$ is the scaled water mass \cite{Walser_1999, Walser_2001, Nguyen_2010, Perkins_2011}. This is an idealised approach with no experimental counterpart that is frequently used to simulate both high and low-viscosity solvents and has been essential for simulation studies investigating internal friction effects \cite{Schulz_2012, DeSancho_2014, Daldrop_2018}. In Fig.~\ref{Figure_Methods}A, we show a snapshot from equilibrium MD simulations of a single united-atom butane molecule suspended in a water-glycerol solvent environment. See Supplementary Information Section S1 for simulation details. The butane dihedral angle $\theta$ (Fig.~\ref{Figure_Methods}B) stochastically transitions between the trans-state, located at $\theta=0$ deg, and the cis-states, located at $\theta=\pm120$ deg. In Fig.~\ref{Figure_Methods}C, we show a typical 0.5 ns trajectory segment for the dihedral of butane in a pure water solvent. The dihedral transitions between states by overcoming barriers in the free-energy landscape. We extract free energy profiles from the trajectory of $\theta(t)$ such that $U(\theta)=-k_{\text{B}}T\text{log}[ \rho(\theta)]$, where $\rho(\theta)$  is the probability density, $T$ is the temperature, and $k_{\text{B}}$ is Boltzmann's constant. In Fig.~\ref{Figure_Methods}D, we show three example free energy profiles extracted from simulations: butane in standard pure neat water, butane in a super-heavy water solvent with ${m}/{m}_0 = 100$, and butane in a water-glycerol solvent with 60$\%$ mass-fraction glycerol. The three free energy profiles are in excellent agreement, indicating that neither viscogenic method affects the equilibrium properties of the dihedral. For the mass-scaled systems, we consider $m/m_0=$ 1, 9, 25, and 100, corresponding to $\eta/\eta_0=$ 1, 3, 5, and 10. Standard pure water for the TIP4P/2005 water model has viscosity $\eta_0=0.86 \; \text{mPa} \text{s}$ \cite{Gonzalez_2010}. For the water-glycerol mixtures, we evaluate viscosity using the Green-Kubo relationship, which relates the shear viscosity to auto-correlations of the shear stress tensor (Supplementary Information Section S2). In Fig.~\ref{Figure_Methods}E, we plot the water-glycerol viscosities for the range of glycerol concentrations used throughout this paper and find excellent agreement compared to an experimental empirical curve for water-glycerol mixtures at $T=300$ K \cite{Cheng_2008}.

\begin{figure*}[t!]
\includegraphics[scale=0.95]{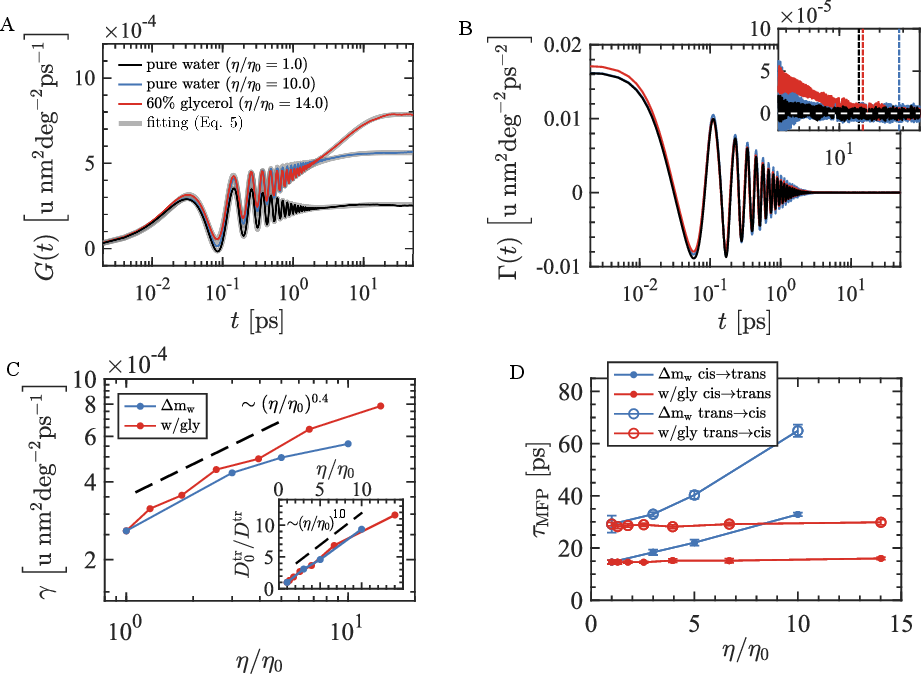}
\caption{(A) Running integrals $G(t)$ extracted from MD trajectories for butane dihedral dynamics under various solvent conditions. The corresponding memory kernels $\Gamma(t)$ are shown in (B) and are evaluated by numerically differentiating $G(t)$. Fits of the running integrals (Eq.~\ref{Fit_Eq}) are underlaid with the corresponding MD curves in (A). The inset in (B) shows the long-time memory kernel decay. The vertical dashed lines indicate the corresponding $\tau_{\rm{MFP}}$. Figure legend in (A) also corresponds to (B). (C) Total friction $\gamma$ acting on the butane dihedral rotation for all super-heavy water and water-glycerol conditions. The black dashed line indicates $\sim(\eta/\eta_0)^{0.4}$ scaling. The inset shows the linear viscosity scaling of the butane centre of mass transition diffusion coefficient $D^{\text{tr}}$, where $D_0^{\text{tr}}$ is the result in standard neat water. (D) Mean first-passage times $\tau_{\text{MFP}}$ for the cis-to-trans (cis$\rightarrow$trans) and trans-to-cis (trans$\rightarrow$cis) transitions of the butane dihedral isomerization.
\label{Figure_Butane_Memory}}
\end{figure*}

To evaluate the friction acting on the butane dihedral, we map $\theta(t)$ onto a generalized Langevin equation (GLE):
\begin{equation}\label{GLE}
\begin{split}
m\ddot{\theta}(t) = -\int\limits_{0}^{t}\Gamma(t-t^{\prime})\dot{\theta}(t^{\prime})dt^{\prime} -\nabla U\big(\theta(t)\big)+  F_R(t),
\end{split}
\end{equation}
where $\Gamma(t)$ is the friction memory kernel, and $F_R(t)$ is the stochastic force term, which has a zero mean $\langle F_R(t) \rangle =0$ and satisfies the fluctuation-dissipation theorem $\langle F_R(t) F_R(t^{\prime}) \rangle =k_{\rm{B}}T\Gamma(t - t^{\prime})$. $U(\theta)$ is the dihedral free energy profile, as given in Fig.~\ref{Figure_Methods}D, and $\nabla \equiv \partial/\partial \theta$. $m$ is the effective mass of the dihedral, which we assume to be independent of $\theta$ since the slight $\theta$-dependence of $m$ has little effect on dihedral dynamics (see Supplementary Information section S3 and \cite{Ayaz_2022}).

We use recent friction memory-kernel extraction techniques and extract the running integral function $G(t) = \int_0^t \Gamma(t^{\prime})dt^{\prime}$ directly from the time series of $\theta(t)$ \cite{Kowalik_2019, Ayaz_2021}. Details of the memory kernel extraction method are given in Supplementary Information Section S4. In Fig.~\ref{Figure_Butane_Memory}A, we show $G(t)$ for the butane dihedral under three solvent conditions. The fitting results for each curve, underlaid in grey, are discussed below. In Fig.~\ref{Figure_Butane_Memory}B, we show the corresponding memory kernels $\Gamma(t)$, evaluated using numerical differentiation of $G(t)$. The large oscillations in $\Gamma (t)$ are due to couplings between the dihedral and bond angle vibrations, the latter being flexible in our model (Supplementary Information Section S5). The inset shows magnifications of the long-time tails, which eventually decay to zero, resulting in the plateauing behaviour of the $G(t)$ functions. To evaluate the total friction $\gamma$ acting on the dihedral, we evaluate $\gamma=G(t\rightarrow\infty)$, given by the plateau values in Fig.~\ref{Figure_Butane_Memory}A. In Fig.~\ref{Figure_Butane_Memory}C, we show $\gamma$ for all solvent conditions as a function of the normalized viscosity $\eta/\eta_0$, where $\eta_0$ is the viscosity of neat water. We see that the viscosity dependence of the friction is almost identical, whether measured in pure, super-heavy water or in the water-glycerol mixtures, and scales approximately as $\gamma(\eta) \sim \eta^{0.4}$, indicating a strong violation of the Stokes-Einstein relation. Interestingly, the translation diffusion coefficients for the butane centre of mass (Fig.~\ref{Figure_Butane_Memory}C inset), which we calculate by fitting the long-time diffusive regime of the mean square displacements (see Supplementary Information Section S6), scale linearly with viscosity in both the water-glycerol mixtures and in the super heavy water, indicating that the translation diffusion does satisfy the Stokes-Einstein relation.

% This result was shown previously for the butane in super-heavy water \cite{Daldrop_2018} and can be attributed to internal friction effects. Need to really take care of this once and for all!!! I think I have the v-rescale results, right?

In Fig.~\ref{Figure_Butane_Memory}D, we show the mean first-passage times $\tau_{\text{MFP}}$ plotted as a function of solvent viscosity for both trans$\rightarrow$cis transitions and cis$\rightarrow$trans transitions. The calculation of $\tau_{\text{MFP}}$ is detailed in Supplementary Information Section S7, where we discuss a method for eliminating recrossing effects. In super-heavy water, $\tau_{\text{MFP}}$ increases significantly with $\eta$. However, in the water-glycerol mixtures, $\tau_{\text{MFP}}$ is completely independent of solvent viscosity. One possibility could be that this is a nano-viscosity effect, where for solutes that are small compared to the size of the viscogenic co-solvent, or similar in size, the viscosity experienced by the solute can deviate away from the measured macroscopic viscosity \cite{Barshtein_1995, Sekhar_2014}. We dismiss this suggestion since both the friction experienced by the rotating dihedral, and the translation diffusion for the butane center of mass, are the same whether measured in the water-glycerol mixtures or in the super-heavy water (Fig.~\ref{Figure_Butane_Memory}C). The results for $\tau_{\text{MFP}}$ indicate that, in the small molecule regime, molecular conformation reaction kinetics can completely decouple from the macroscopic viscosity of the solvent environment. In fact, the disparities between $\tau_{\text{MFP}}(\eta)$ measured in super-heavy water and the water-glycerol mixtures result from complex non-Markovian effects, which we return to below. \\

\begin{figure*}[!t]
\includegraphics[scale=0.92]{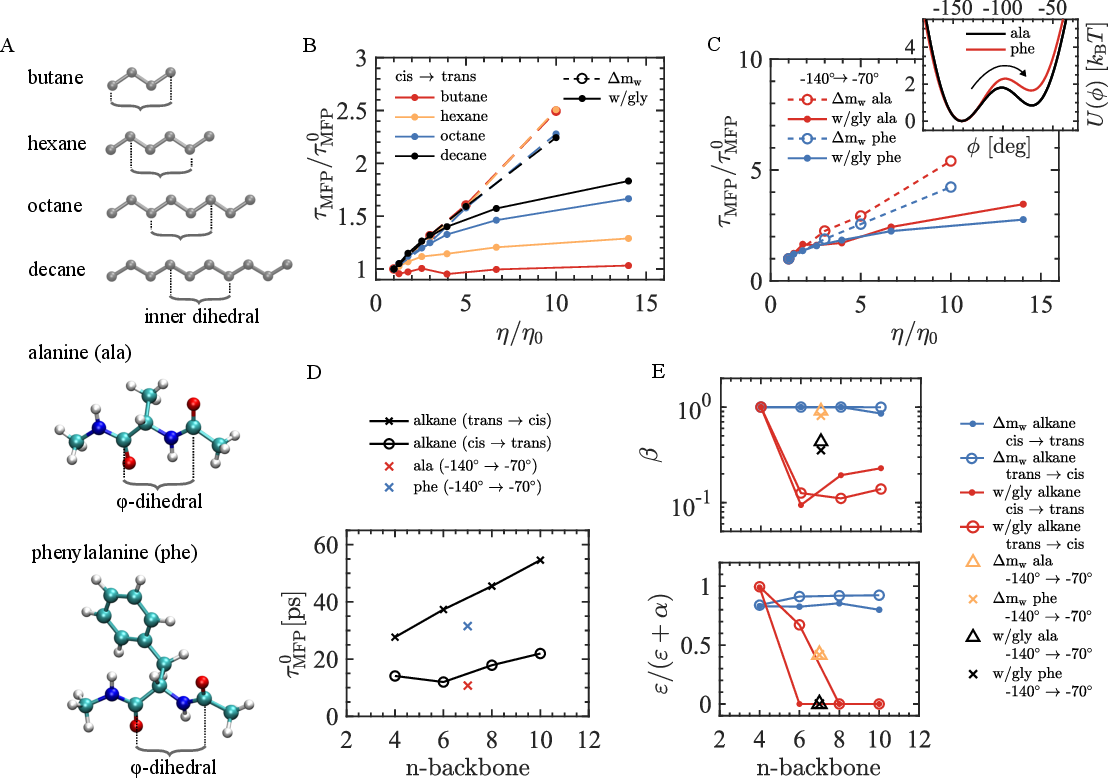}
\caption{Molecular-size dependence of isomerization kinetics. (A) Molecular schematic for alkanes and two capped, dipeptide-bond amino acids. For the alkanes, the inner dihedrals are indicated. For alanine (\textit{ala}) and phenylalanine (\textit{phe}), the $\phi$-dihedrals are indicated. (B) Cis-to-trans mean first-passage times $\tau_{\text{MFP}}$ for alkanes. $\tau_{\text{MFP}}$ values are scaled by the reaction times measured in neat water (pure water at standard-viscosity) for each system $\tau^0_{\text{MFP}}$. The left legend indicates the colour scheme for the different alkanes. The right legend indicates the solvent type. (C) Scaled reaction times for amino acids. Reaction times are shown for the $-140^{\circ}$ to $-70^{\circ}$ transition, as indicated by the arrow on the free-energy profiles (inset). (D) Reaction times in neat water $\tau^0_{\text{MFP}}$. Reaction times for alkanes are plotted as a function of backbone length (number of carbon atoms). For the amino acids, the backbone length is $n=7$. (E) Fitting parameters $\beta$ and $\varepsilon/(\alpha+\varepsilon)$ for scaling relations: $\tau_{\text{MFP}}(\eta) = \alpha(\eta/\eta_0)^{\beta} + \varepsilon$ (See Supplementary Information Section S9).  \label{n_Alkanes}}.
\end{figure*}

\noindent \textbf{Molecular-size dependence of isomerisation kinetics. }To show that the decoupling between $\tau_{\text{MFP}}$ and $\eta$ in the water-glycerol solvent is a small-molecule effect, we systematically increase the length of the alkane chain to include hexane, octane, and decane. We then evaluate the mean first-passage times for the inner-most dihedral of each chain (Fig.~\ref{n_Alkanes}A, see supplementary Information Section S8 for details). Additionally, we simulate capped alanine and phenylalanine amino acids and evaluate $\tau_{\text{MFP}}$ for the $\phi$-dihedral. The mean first-passage times for the alkanes exhibit a clear solute-size dependence (Fig.~\ref{n_Alkanes}B). Specifically, when rescaled by $\tau^0_{\text{MFP}}$ (the result for neat water), the isomerization times for both octane and decane show convergent scaling between the super-heavy water and water-glycerol mixtures in the low viscosity regime. The results for alanine and phenylalanine (Fig.~\ref{n_Alkanes}C) are consistent with the alkane results since the backbone lengths of the two amino acids are both seven heavy atoms long, between that of hexane and octane. De Sancho et al. show that the viscosity scaling of isomerization kinetics in the super-heavy water solvent is essentially the same for a range of dipeptides and that there are only slight deviations for the alanine dipeptide in the super-heavy water solvent when compared to a mixed glucose-water solvent \cite{DeSancho_2014}. However, they only measure in the range of  $1< \eta/\eta_0 < 3.5$, where they interpret the difference as negligible, suggesting that the scaling is therefore identical in the two viscogens (water + glucose and super-heavy water). By extending the viscosity range, our investigation reveals that slight deviations are present and that the deviations increase with increasing viscosity.  

In Fig.~\ref{n_Alkanes}D, we show the dihedral isomerization times in neat water $\tau^0_{\text{MFP}}$. The -140$^{\circ}\rightarrow$ -70$^{\circ}$ transition in phenylalanine $\tau^0_{\text{MFP}}$ is much greater than that of alanine. This increase is only partially due to the 20$\%$ increase in the phenylalanine barrier height (Fig.~\ref{n_Alkanes}C inset and Supplementary Information Section S8) with the remaining contribution coming from the presence of the large benzyl side group. In Fig.~\ref{n_Alkanes}C, we see that the viscosity scaling for the two dipeptides is similar, suggesting that the addition of the large benzyl side group does not significantly affect the viscosity scaling of the isomerization times, but rather just the absolute values. De Sancho et al. also addressed the issue of size dependence by expanding the radius of the united-atom groups in their n-butane model, which led to longer mean isomerization times \cite{DeSancho_2014}. Our results demonstrate that varying the width of a molecule has a different effect compared to increasing a dihedral chain length, where it is the length of a dihedral backbone that predominantly influences isomerization rate scaling.

\begin{figure*}[!t]
\includegraphics[scale=0.95]{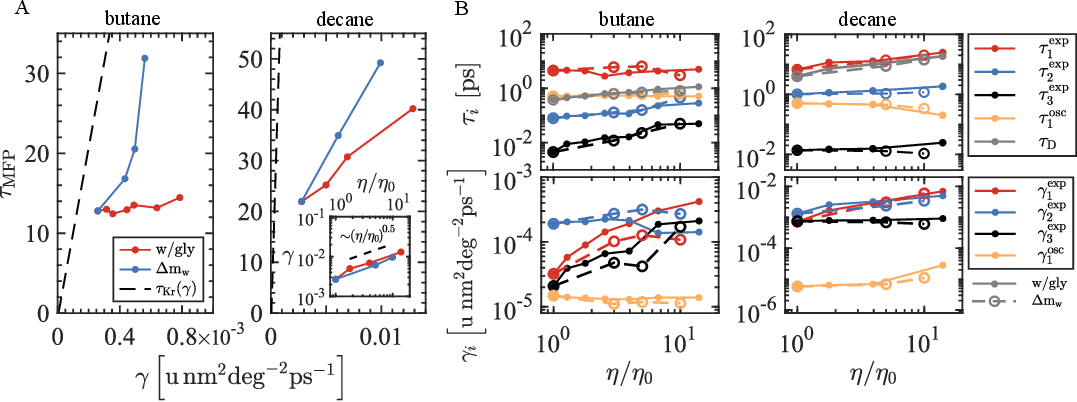}
\caption{(A) Dependence of mean first-passage time $\tau_{\text{MFP}}$ on dihedral friction for butane and decane. $\tau_{\text{Kr}}(\gamma)$ is the overdamped, high friction Kramers prediction (black dashed line). The inset for decane shows the friction-viscosity scaling, which scales as $\gamma(\eta)\sim (\eta/\eta_0)^{0.5}$. For butane friction-viscosity scaling, see Fig.~\ref{Figure_Butane_Memory}C. (B) Fitting parameters for Eq.~\ref{Fit_Eq} for fits of the memory kernels of butane and the inner dihedral of decane. $\gamma_i$ is for the set of normalized coefficients $\gamma_1^{\rm{exp}}$, $\gamma_2^{\rm{exp}}$, $\gamma_3^{\rm{exp}}$, and $\gamma_1^{\rm{osc}}$, written such that $\gamma_1^{\rm{exp}}+\gamma_2^{\rm{exp}}+\gamma_3^{\rm{exp}}+\gamma_1^{\rm{osc}} = \gamma$, where $\gamma$ is the total friction on the dihedral, and $\tau_i$ is for the set of corresponding time scales. The time scales are compared to the diffusion times $\tau_{\text{D}}= \gamma L^2/k_{\text{B}}T$, where $L=60\;\text{deg}$. \label{Fig_4}}.
\end{figure*}

We investigate the scaling relations shown in Figs.~\ref{n_Alkanes}B and C in more detail by fitting $\tau_{\rm{MFP}}$ for each molecule with the scaling function 
\begin{equation}\label{scaling}
\tau_{\text{MFP}}(\eta) = \alpha\bigg( \frac{\eta}{\eta_0}\bigg)^{\beta} + \varepsilon
\end{equation}
(see Supplementary Information Section S9 for details). There are two ways that Eq.~\ref{scaling} can reveal strong internal friction effects. The first is when $\varepsilon/(\alpha + \varepsilon) \rightarrow 1$, indicating that contributions from the non-zero intercept ($\varepsilon>0$) dominate the pre-factor $\alpha$. The second is when $\beta<1$, such that the scaling of Eq.~\ref{scaling} is sublinear. In Fig.~\ref{n_Alkanes}E, we show the exponent $\beta$ and the ratio $\varepsilon/(\alpha + \varepsilon)$ for all systems. For alkanes in super-heavy water (blue data), the scaling is effectively independent of chain length, with $\beta\approx 1$ indicating linear scaling for all systems. However, for all alkanes, $\varepsilon/(\alpha + \varepsilon)$ is between 0.8 and 0.9, indicating that despite the linear scaling, strong internal friction effects are present. For butane, it was previously shown that $\tau_{\text{MFP}}$ scales linearly with viscosity for $\eta>\eta_0$ but that the scaling transitions to an inertia-dominated regime for $\eta<\eta_0$ \cite{Daldrop_2018}. We do not consider this regime here. However, it is interesting to note that the approximate linear scaling for $\eta/\eta_0>1$ is a characteristic of all alkanes in super-heavy water. For alkanes in water-glycerol solvents (red data), $\varepsilon/(\alpha + \varepsilon)$ transitions from 1 to 0 for longer alkane chains, accompanied by a significant decrease in $\beta$. Here, butane is insensitive to changes in viscosity such that $\beta\rightarrow 1$ and $\alpha \rightarrow 0$ represent a constant function. However, chains longer than butane are sensitive to $\eta$ such that $\tau_{\text{MFP}}(\eta) = \alpha(\eta/\eta_0)^{\beta}$, with $\varepsilon=0$ and sub-linear scaling such that $\beta < 0.25$. The dipeptides are interesting because they also scale linearly in the super-heavy water but with relatively reduced internal friction contributions ($\varepsilon/(\alpha + \varepsilon)\approx 0.4$). In contrast, in the water-glycerol solvents, the peptides are described by $\varepsilon=0$ but with $\beta \approx 0.45$. Altogether, these results confirm two distinct behaviours for the influence of internal friction effects on the viscosity dependence of dihedral isomerization kinetics and that it is not just the construction of the molecule, or the viscosity of the solvent that determines the nature of the viscosity scaling, but also the composition of the solvent. \\
 
\noindent \textbf{Memory-induced speed-up of isomerization kinetics. } In Fig.~\ref{Fig_4}A, we show the dependence of the cis-to-trans mean first-passage times $\tau_{\text{MFP}}$ for the butane dihedral and the inner decane dihedral on the extracted dihedral friction. For decane, we also show the friction-viscosity scaling (inset), which, like butane (Fig.~\ref{Figure_Butane_Memory}C), is strongly sublinear. In the high friction, overdamped limit, Kramers' theory predicts that $\tau_{\text{Kr}}(\gamma) = 2\pi\gamma \text{e}^{U_0/k_{\text{B}}T}/\sqrt{|U^{\prime\prime}_{\text{min}}U^{\prime\prime}_{\text{max}}|}$. From Fig.~S6, the free energy profiles for all alkanes are approximately the same. $U^{\prime\prime}_{\rm{max}} = -1.35{\times}10^{-3}$ $k_{\text{B}}T/\text{deg}^2$ and $U^{\prime\prime}_{\rm{min}} = 6.25{\times}10^{-3}$ $k_{\text{B}}T/\text{deg}^2$ are the curvatures of the free energy at the maximum and minimum, and $U_0 = 3.9$ $k_{\text{B}}T$ is the free energy barrier height. For butane, $\tau_{\text{Kr}}(\gamma) $ clearly does not represent the water-glycerol system well (Fig.~\ref{Fig_4}A), and it overestimates the super-heavy water results by as much as a factor of 3. For decane, the range of $\gamma$ is an order of magnitude higher than butane. However, the Kramers pre-factor is the same such that the Kramers prediction $\tau_{\text{Kr}}(\gamma)$ for decane dramatically overestimates the measured values by a factor of between 13 and 25. In the Supplementary Information Section S10, we also compare predictions for the Grote-Hynes theory \cite{Grote_1980}, which explicitly accounts for frequency-dependent friction effects, and the more general intermediate-friction Kramers' theory \cite{Kramers_1940}, to the measured MD reaction times. Neither theory consistently predicts the reaction times for butane and decane isomerization in both the water-glycerol solvent and the super heavy water. The comparison between the MD data and the Kramers predictions in Fig.~\ref{Fig_4}A suggest that memory-induced barrier crossing speed-up effects are present. 

To quantify memory effects, we fit the following function to the memory kernels for the butane dihedral and the inner decane dihedral \cite{Brunig_2022a}:
\begin{equation}\label{Fit_Eq}
\begin{split}
&\Gamma(t) \approx\sum_{i=1}^3\frac{\gamma_i^{\rm{exp}}}{\tau_i^{\rm{exp}}}{e}^{-t/\tau_i^{\rm{exp}}} \\
& \; +\frac{(1+\omega_1 \tau_1^{\rm{osc}})\gamma_1^{\rm{osc}}}{2\tau_1^{\rm{osc}}} {e}^{-t/\tau_1^{\rm{osc}}}\bigg[ \text{cos}( \omega_1 t ) + \frac{\text{sin}( \omega_1 t)}{\omega_1 \tau^{\text{osc}}}  \bigg].
\end{split}
\end{equation} 
Here, $\gamma_i^{\rm{exp}}$ and $\tau_i^{\rm{exp}}$ are the amplitudes and time scales for exponentially decaying modes. $\gamma_1^{\rm{osc}}$ is the amplitude, $\omega_1$ is the angular frequency, and $\tau_1^{\rm{osc}}$ is the decay time for the decaying-oscillating term. Eq.~\ref{Fit_Eq} is written such that $\gamma_1^{\rm{exp}}+\gamma_2^{\rm{exp}}+\gamma_3^{\rm{exp}}+\gamma_1^{\rm{osc}} = \gamma$. Fits of Eq.~\ref{Fit_Eq} are shown in Fig.~\ref{Figure_Butane_Memory}A and Figs.~S13 and S14 of Supplementary Information Section S10. It has been shown that for systems with multi-exponential memory kernels, memory components with times scales much longer than the diffusion time $\tau_{\text{D}}$ do not affect mean first-passage times \cite{Kappler_2019, Lavacchi_2020}. We evaluate the diffusion times using $\tau_{\text{D}}=\gamma L^2/k_{\text{B}}T$, where $L$ represents a characteristic length on the reaction coordinate (here taken to be 60 deg - the distance from the cis minimum to the barrier top). For butane, the longest exponential time scale $\tau_1^{\text{exp}}$ is much greater than $\tau_{\text{D}}$ (Fig.~\ref{Fig_4}B). Therefore, the intermediate exponential mode, for which $\tau_2^{\text{exp}} < \tau_{\text{D}}$, has the dominant influence on $\tau_{\text{MFP}}$. This is the only mode for which the amplitudes ($\gamma_2^{\text{exp}}$) are greater in the super-heavy water than in the water-glycerol mixtures, which helps to explain the divergent behaviour for $\tau_{\text{MFP}}$ in Fig.~\ref{Figure_Butane_Memory}D. Although the decaying time scale for the oscillating mode $\tau_1^{\text{osc}}$ is approximately equal to $\tau_{\text{D}}$, the amplitudes $\gamma_1^{\text{osc}}$ are relatively small. (In Supplementary Information section S11, we investigate the influence of the oscillating contributions to the Grote-Hynes prediction and show no effect on the viscosity scaling of the barrier crossing times.) For decane, the longest exponential mode is likely the dominant contribution since $\tau_1^{\text{exp}} \approx \tau_{\text{D}}$. This time scale analysis is consistent with a recent analysis of extensive protein folding simulations \cite{Dalton_2023}, where it was shown that proteins fold in a memory-induced speed-up regime. Decane exhibits strongly accelerate kinetics (Fig.~\ref{Fig_4}A), which is expected for systems with $\tau_1^{\text{exp}} \approx \tau_{\text{D}}$ \cite{Kappler_2019, Lavacchi_2020} since such systems are most sensitive to kinetic speed-up effects. As memory time scales exceed $\tau_{\text{D}}$, systems approach a memory induced slow-down regime, which has been shown to be the kinetic mode for some proteins \cite{Dalton_2023}. For butane, speed-up effects are present but weaker, suggesting that butane is closer to the memory induced slow-down transition, which is consistent with the measurement that $\tau_1^{\text{exp}} > \tau_{\text{D}}$. Overall, these results reveal that dihedral isomerization in viscous solvents exhibits multi-time-scale non-Markovian dynamics with memory-accelerated isomerization kinetics.

 % Overall, our results are consistent with previous investigations of non-Markovian reaction kinetics for molecular reconfiguration dynamics in solvents confirming that the coupling between small molecule isomerization dynamics and the viscosity of  molecular solvents is a multi-time-scale non-Markovian problem.

\section*{Discussion and conclusions}

We utilize recent memory kernel extraction techniques to directly evaluate the frequency-dependent friction acting on an isomerizing dihedral. In doing so, we explore the relationship between the friction acting on a dihedral, the viscosity of a solvent, and molecular reconfiguration kinetics, and we do so for a variety of molecular solutes in different solvent conditions. Our study reveals two significant findings. Firstly, the total butane isomerization friction $\gamma$ scales equivalently with viscosity, whether measured in a water-glycerol mixture or the super heavy water (Fig.~\ref{Figure_Butane_Memory}C). In both scenarios, this scaling strongly deviates from the linear scaling expected according to the Stokes-Einstein relation. Secondly, the isomerization kinetics differ markedly when measured in a water-glycerol mixture compared to super heavy water. For butane, the mean first-passage times become completely decoupled from viscosity in the water-glycerol mixture but scale linearly in the super-heavy water (Fig.~\ref{Figure_Butane_Memory}D). We can exclude nano-viscosity effects as the cause of this difference since both the translational diffusion coefficient for butane and the friction-viscosity scaling of the dihedral rotation are equivalent in both the water-glycerol mixture and super-heavy water (Fig.~\ref{Figure_Butane_Memory}C). We suggest that this difference in kinetic behaviour between the two solvation methods arises from the multi-time-scale nature of the frequency-dependent friction. Different time-scale contributions of friction interact in distinct ways with the mixed water-glycerol solvent and the super-heavy water. We confirm the significance of non-Markovian contributions by demonstrating that dihedral isomerization times are much shorter than the predictions of Kramers' theory in the high-friction, overdamped limit, which is due to memory-induced acceleration effects.

The viscosity-dependent isomerization times of larger molecules, such as decane or capped amino acids, appear to converge between the two solvation methods, at least in the lower viscosity regime (Figs.~\ref{n_Alkanes}B and C). These results could be validated experimentally. Evidence suggests that the viscosity scaling of relaxation rates remains largely consistent for hairpin-forming polypeptide chains, regardless of whether they are dissolved in a glucose or sucrose co-solvent.  However, deviations have been observed for helix-forming polypeptide chains \cite{Jas_2001}. Similarly, Sekhar et al. showed that the interconversion rates of a four-helix bundle domain are different when measured in either a mixed water-glycerol solvent or a mixed water-bovine serum albumin (BSA) solvent \cite{Sekhar_2014}, which they attributed to micro-viscosity effects. However, as we have shown here, these difference may be rather due to complex non-Markovian effects that result from the interactions between the protein domain and mixed solvent environments.  Another example where the current investigation is directly applicable is in the study of molecular rotor dyes, where the reconfiguration kinetics of a dye molecule are used to estimate the viscosity of some complex viscous environment \cite{Forster_1971, Grabowski_2003, Akers_2004, Haidekker_2010a, Haidekke_2010}. These fluorescent molecules undergo stochastic isomeric switching at rates determined by the viscosity of their environment. Currently, it remains uncertain to what degree multi-time-scale friction effects are important for these dyes, especially in complex viscogenic environments. Overall, there are many interesting areas where accurate measurements of friction, viscosity, and reaction kinetics are essential for understanding molecular processes and complex solvent-solute coupling.

\section*{Methods} For further information on simulation details, see Supplementary Information Section S1. Additional details regarding the simulations and analysis, including the evaluation of solvent viscosities, extraction of friction memory kernels, and various fitting procedures, are also available in the Supplementary Information document. \\
\section*{Acknowledgements} The project was supported by the European Research Council (ERC) Advanced Grant 835117 NoMaMemo and the Deutsche Forschungsgemeinschaft (DFG) Grant No. SFB 1449 "Dynamic Hydrogels at Biointerfaces". The authors would like to acknowledge the HPC Service of ZEDAT, Freie Universität Berlin, for providing computing time. We are also thankful to the physics-department HPC services at Freie University of Berlin for their generous support.

% \bibliography{/Users/dalton/Documents/Mendeley_Bibtex/Freie_Uni.bib}
\bibliography{bib_file.bib}

\end{document}